\documentclass[10pt]{iopart}
\usepackage{graphicx}
\usepackage{setstack}
\usepackage{amssymb}

\begin{document}
\title[Exchange interactions and Curie temperatures in Mn$_2$CoZ compounds]{Exchange interactions and Curie temperatures in Mn$_2$CoZ compounds}
\author{Markus Meinert, Jan-Michael Schmalhorst, and G\"unter Reiss}
\address{Department of Physics, Bielefeld University,
33501 Bielefeld, Germany}
\ead{meinert@physik.uni-bielefeld.de}

\date{\today}

\begin{abstract}
The generalized Heusler compounds Mn$_2$Co\textit{Z} (\textit{Z} = Al, Ga, In, Si, Ge, Sn, Sb) with the Hg$_2$CuTi structure are of large interest due to their half-metallic ferrimagnetism. The complex magnetic interactions between the constituents are studied by first principles calculations of the Heisenberg exchange coupling parameters, and Curie temperatures are calculated from those. Due to the direct Mn-Mn exchange interaction in Mn$_2$Co\textit{Z}, the Curie temperature \textit{decreases}, while the total moment \textit{increases} when changing \textit{Z} from one group to another. The exchange interactions are dominated by a strong direct exchange between Co and its nearest neighbor Mn on the B site, which is nearly constant. The coupling between the nearest-neighbor Mn atoms scales with the magnetic moment of the Mn atom on the C site. Calculations with different lattice parameters suggest a negative pressure dependence of the Curie temperature, which follows from decreasing magnetic moments. Curie temperatures of more than 800\,K are predicted for Mn$_2$CoAl (890\,K), Mn$_2$CoGa (886\,K), and Mn$_2$CoIn (845\,K).

\end{abstract}

\maketitle

\section{Introduction}
Recently, the Mn$_2$\textit{YZ} compounds with the Hg$_2$CuTi structure have attracted considerable theoretical and experimental activities, where \textit{Y} = Fe, Co, Ni, Cu \cite{Luo08_1, Liu08, Xing08, Lakshmi05, Dai06, Helmholdt87, Luo09_1, Luo09_2, Wei10, Li09}. Some of these compounds have been characterized as half-metallic ferrimagnets. The Hg$_2$CuTi structure is closely related to the structure of the well known Heusler compounds. The Mn$_2$\textit{YZ} compounds also follow the Slater-Pauling rule connecting the magnetic moment $m$ and the number of valence electrons $N_V$ via $m = N_V - 24$ in half-metallic Heusler compounds \cite{Galanakis02}.

Half-metallic compounds are characterized by a gap for either the spin-down or the spin-up density of states (DOS) at the Fermi energy, so that an electric current has purely up or down electrons. In applications that make use of spin polarized currents, materials with this property are obviously favorable. The concept was discovered by de Groot \textit{et al.} in band structure calculations of the semi-Heusler compound NiMnSb \cite{deGroot83}. Since then, many more systems were characterized as half-metallic by band structure theory, in particular from the class of Heusler compounds.

Heusler compounds have the general chemical formula \textit{X}$_2$\textit{YZ}, where \textit{X} and \textit{Y} are transition metals, and \textit{Z} is a main group III, IV, or V element and adopt the L2$_1$ crystal structure (space group Fm$\bar{3}$m). It can be imagined as four interpenetrating fcc lattices with the basis vectors $A = (0, 0, 0)$, $B = (1/4, 1/4, 1/4)$, $C = (1/2, 1/2, 1/2)$, and $D = (3/4, 3/4, 3/4)$. The sites $B$ and $D$ are equivalent by symmetry and occupied by the \textit{X} element, whereas \textit{Y} and \textit{Z} occupy the $A$ and $C$ sites. In the Hg$_2$CuTi structure, the $B$ and $C$ sites are occupied by \textit{X}, which makes them nearest neighbors. This occupation is preferred with repect to the Heusler structure if \textit{X} has less valence electrons than \textit{Y} \cite{Luo08_0, Luo09_1}. In this structure, the inversion symmetry is broken and it is thus assigned to space group F$\bar{4}$3m. It can be thought of as a generalization of the Heusler structure.

Half-metallic ferrimagnets have advantages over the well-known half-metallic ferromagnets: due to the internal spin compensation they have low magnetic moment and weak stray field, while the Curie temperature can be high \cite{Pickett01}. The probably best known half-metallic ferrimagnet among the Heusler compounds is Mn$_2$VAl \cite{Itoh83,Yoshida81,Jiang01,Ishida84,Weht99}. It has been predicted to be a half-metallic ferrimagnet \cite{Oezdogan06,Sasioglu05_1} with a low magnetic moment of $2\,\mu_\mathrm{B}$ per formula unit (f.u.). The Curie temperature of 760\,K is the highest reported to date for the Mn$_2$ based (generalized) Heusler compounds and makes Mn$_2$VAl a promising compound for spintronics \cite{Jiang01}. Apart from that, several other materials classes have been proposed to be half-metallic ferrimagnets \cite{Nakamura05, Galanakis06}.

Ideally, an electrode material for spintronics would be a half-metal with tunable net moment. For net moment of zero, this would be a half-metallic antiferromagnet or fully compensated half-metallic ferrimagnet, and several materials have been predicted to show this unusual property \cite{vanLeuken95, Pickett98, Akai06, Galanakis07_1, Oezdogan09, Galanakis07_2}. However, half-metallic antiferromagnetism is limited to a small temperature range because of the inequivalent magnetic sublattices \cite{Sasioglu09}.

In the literature it has been noted that the Mn$_2$\textit{YZ} compounds with Hg$_2$CuTi structure are dominated by direct exchange between the nearest neighbor Mn atoms, but direct calculations of the exchange interactions are missing. It is the scope of this paper to provide these calculations for the Mn$_2$Co\textit{Z} compounds. We focus on this compound series because it has been experimentally synthesized, and band structure calculations suggested very large atomic moments and half-metallicity in most cases.

The half-metallicity of Mn$_2$Co\textit{Z} is constituted by two processes \cite{Liu08}. First, a broad covalent gap of Mn(B) is created by covalent hybridization with Co and Mn(C), which form the (double tetrahedral) nearest neighbor shell. However, the final size of the minority gap is determined by the $e_u$-$t_{1u}$ splitting in the hybridization of Co and Mn(C), which form each other's (octahedral) second nearest neighbor shells. Mn(B) states do not contribute to this hybridization because of the different symmetry transformations. Thus, the band gap is a \textit{d-d} gap \cite{Fang02}. This situation is similar to the one in the Co$_2$Mn\textit{Z} Heusler compounds, where the $e_u$-$t_{1u}$ splitting of the Co-Co hybridization governs the minority gap \cite{Galanakis02}.

\section{Computational approach}

We performed our calculations with the spin-polarized relativistic Korringa-Kohn-Rostoker package \textit{Munich} SPRKKR \cite{sprkkr}. The calculations were carried out in the full-potential mode with an angular momentum cutoff of $l_\mathrm{max} = 3$ on a $28 \times 28 \times 28$ $\bi{k}$ point mesh (564 points in the irreducible wedge of the Brillouin zone). In order to further improve the charge convergence with respect to $l_\mathrm{max}$, we employed Lloyd's formula for the determination of the Fermi energy \cite{Lloyd72, Zeller08}. The exchange-correlation potential was modeled within the generalized gradient approximation of Perdew, Burke, and Ernzerhof \cite{PBE} and the calculations were converged to about 0.1 meV. All calculations were carried out in the scalar-relativistic representation of the valence states, thus neglecting the spin-orbit coupling.

In the classical Heisenberg model the Hamiltonian of a spin system is given by
\begin{equation}\label{eq:heisenberg}
H = -\sum\limits_{i,j} \bi{e}_i \bi{e}_j J_{ij},
\end{equation}
with the Heisenberg pair exchange coupling parameters $J_{ij}$, and unit vectors $\bi{e}_{i}$ pointing in the direction of the magnetic moment on site $i$. SPRKKR allows to calculate the exchange coupling parameters by mapping the system onto a Heisenberg Hamiltonian. The parameters are determined within a real-space approach using the theory by Liechtenstein \textit{et al.} \cite{Liechtenstein87}. From the $J_{ij}$ the Curie temperatures were calculated within the mean field approximation (MFA). For a single-lattice system the Curie temperature is given within the MFA by
\begin{equation}\label{eq:single}
\frac{3}{2} k_\mathrm{B} T_\mathrm{C}^{\mathrm{MFA}} = J_0 = \sum\limits_j J_{0j}.
\end{equation}
In a multi-sublattice system---as, e.g., the Heusler compounds with four sublattices---one has to solve the coupled equations
\begin{eqnarray}\label{eq:multi}
\frac{3}{2} k_\mathrm{B} T_\mathrm{C}^{\mathrm{MFA}} \left< e^\mu \right> &=& \sum\limits_\nu J_0^{\mu \nu} \left< e^\nu \right>\\
J_0^{\mu \nu} &=& \sum\limits_{\bi{r}\neq 0} J^{\mu \nu}_{0 \bi{r}}, \nonumber
\end{eqnarray}
where $\left< e^\nu \right>$ is the average $z$ component of the unit vector $e_\bi{r}^\nu$ pointing in the direction of the magnetic moment at site ($\nu$, $\bi{r}$). The coupled equations can be rewritten as an eigenvalue problem:
\begin{eqnarray}\label{eq:eigen}
(\bi{\Theta} - T\, \bi{I})\,\bi{E} &=& 0\\
\frac{3}{2} k_\mathrm{B} \Theta_{\mu \nu} &=& J_0^{\mu \nu}
\nonumber
\end{eqnarray}
with a unit matrix $\bi{I}$ and the vector $E^\nu = \left< e^\nu \right>$. The largest eigenvalue of the $\bi{\Theta}$ matrix gives the Curie temperature \cite{Sasioglu05_1, Anderson63}. The $\bi{r}$-summation in Eq. (\ref{eq:multi}) was taken to a radius of $r_\mathrm{max} = 3.0\,a$, where $a$ is the lattice constant.

The lattice parameters were taken from Liu \textit{et al.} \cite{Liu08}, who provide experimental values for $Z$ = Al, Ga, In, Ge, Sn, Sb. For Mn$_2$CoSi we assumed the Mn$_2$CoGe parameter reduced by 0.1\,\AA{}, which is observed, e.g., for Co$_2$MnSi -- Co$_2$MnGe \cite{Webster71}. The calculations of Mn$_2$CoIn and Mn$_2$CoSn were unstable at the experimental lattice parameters, but could be stabilized with slightly reduced values. All lattice parameters used here are summarized in Table \ref{overview}.

\section{Results}

\Table{\label{overview} Lattice parameters used for the calculations and resulting total and site resolved magnetic moments. The total magnetic moments are given in $\mu_B$ per formula unit, the atomic magnetic moments are given in $\mu_B$ per atom.}
\br
Mn$_2$Co\textit{Z}	& $a$ (\AA{})	 	&	$m_\mathrm{total}$		& $m_\mathrm{Co}$	& $m_\mathrm{Mn(B)}$	& $m_\mathrm{Mn(C)}$ 	& $m_Z$\\ \mr
Al	& 5.84		&	1.99		& 0.94		& 2.69		&	-1.59	&	-0.05\\
Ga	& 5.86		&	2.01		& 0.93		& 2.88		&	-1.78	&	-0.03\\
In	& 6.04$^{a}$	&	1.95		& 0.99		& 3.16		&	-2.18	&	-0.02\\ \bs

Si	& 5.70	&	2.99		& 0.84		& 2.66		&	-0.50	&	-0.01\\
Ge	& 5.80		&	2.98		& 0.87		& 2.83		&	-0.72	&	0.01\\
Sn	& 5.96$^{a}$	&	2.98		& 0.83		& 2.96		&	-0.81	&	-0.01\\ \bs

Sb	& 5.90		&	3.97		& 0.88		& 2.95	 	&	0.15	&	0.00\\
\br
\end{tabular}
\item[$^{a}$] exp. lattice parameters: Mn$_2$CoIn 6.14\,\AA{}, Mn$_2$CoSn 6.06\,\AA{}
\end{indented}
\end{table}

\subsection{Magnetic moments and densities of states}

The electronic structure calculations yield a half-metallic ground state in all cases with the exception of Mn$_2$CoGa and Mn$_2$CoIn. Our results for the total and site resolved magnetic moments are summarized in Table \ref{overview}. The total magnetic moments closely follow the Slater-Pauling rule for half-metallic Heusler compounds, so that we have magnetic moments of 2, 3, or 4\,$\mu_B$\,/\,f.u. if \textit{Z} is a group III, IV, or V element, respectively. Small deviations from the integer values arise from the angular momentum truncation at $l_\mathrm{max}=3$, which gives rise to a very small DOS in the minority gap. This is a typical observation when using the KKR method on ferromagnetic half-metals (see, e.g., Galanakis \textit{et al.} \cite{Galanakis02}). The magnetic moment of the Co atom is nearly constant for different \textit{Z} materials, being about 0.9\,$\mu_B$. Similarly, the Mn(B) atom has a nearly constant magnetic moment in the range of 2.69 to 3.16\,$\mu_B$. In contrast, the moment of the Mn(C) atom changes considerably with the valence electron number and determines finally the total moment. All Mn$_2$Co\textit{Z} compounds are ferrimagnetic due to the Mn(C) atom with the exception of Mn$_2$CoSb, which is a ferromagnet. In all cases the \textit{Z} atom is nearly unpolarized. Only small changes are observed for the site resolved moments when \textit{Z} is changed within on group. The increase of the absolute value of the Mn moments can be traced to the lattice parameter change upon \textit{Z} change. The orbital overlap is reduced with increasing lattice parameter, giving rise to weaker hybridizations (which is also the reason for the gap width reduction). Because of this reduction of itinerancy the quenching of the atomic moments is less effective and the moments become more atomic-like, i.e., larger.

Our results differ considerably from those given by Liu \textit{et al.} \cite{Liu08}, who used the full potential linearized augmented plane waves (FLAPW) method. The total magnetic moments are in very good agreement, but the atomic moments are smaller in our calculations by 0.3 to 0.7\,$\mu_B$ for Mn(B) and Mn(C). In contrast, the magnetic moments of the Co atoms are nearly equal. Most notably, in our calculations Mn$_2$CoSb is ferro- instead of ferrimagnetic. Therefore, we have checked our SPRKKR results with the FLAPW package Elk \cite{elk}. Our FLAPW results are concordant with the SPRKKR data, leaving the discrepancies with Liu \textit{et al.} unexplained.

Apart from these differences the DOS are in good agreement with \cite{Liu08} and all conclusions about the electronic structure given there are transferable to our calculations.

\subsection{Exchange interactions and Curie temperatures}

\begin{figure}[t]
\centering\includegraphics[scale=1]{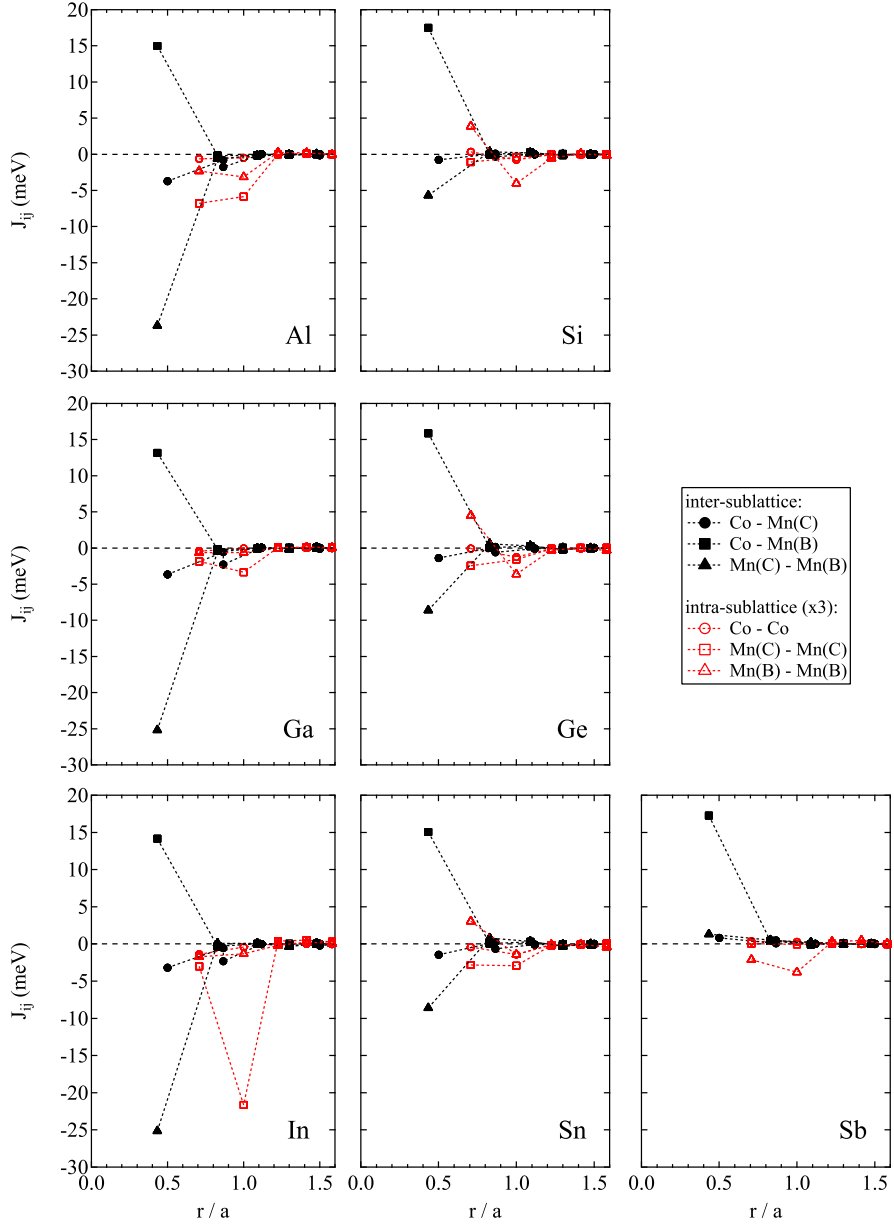}
\caption{\label{fig:jxc}Heisenberg exchange coupling parameters $J_{ij}$ for the Mn$_2$Co\textit{Z} compounds as a function of the interatomic distance $r$. Note that the intra-sublattice interactions have been multiplied by 3 for clarity.}
\end{figure}

Figure \ref{fig:jxc} shows the Heisenberg exchange coupling parameters obtained from our calculations. To ease the following discussion, refer to Table \ref{coordinations} for the atomic coordinations.

\Table{\label{coordinations} Nearest and second nearest neighbor coordinations in Mn$_2$Co\textit{Z}.}
\br
distance (r/a)		& 0.433			&	&	0.50		\\
symmetry		& $T_d$			&	&	$O_h$		\\ \mr
Co			& Mn(B) / \textit{Z}	& 	&	Mn(C)		\\
Mn(B)			& Mn(C) / Co		&	&	\textit{Z}		\\
Mn(C)			& Mn(B) / \textit{Z}	&	&	Co		\\
\textit{Z}		& Mn(C) / Co		&	&	Mn(B)		\\
\br
\end{tabular}
\end{indented}
\end{table}

We start with the discussion of the Al--Ga--In series. It is notable that the exchange interactions are tightly confined to clusters of radius $r \leq a$. In particular, the inter-sublattice interactions have significant contributions only for the nearest and second nearest neighbors, while the intra-sublattice contributions are significant up to $r=a$. An exponential damping of the exchange interactions is expected for half-metals \cite{Rusz06}; in the cases of Ga  and In the damping is also present, but not as efficient as in the half-metallic case of Al. One observes clearly the dominating Co-Mn(B) and Mn(C)-Mn(B) nearest neighbor interactions, where the Mn(C)-Mn(B) interaction is clearly the stronger one. The Co-Mn(B) (second nearest neighbor interaction) is much weaker in comparison. In the graphs we omit the interactions with \textit{Z}, because these are effectively zero for all distances. Co and Mn(C) couple antiferromagentically to Mn(B), while Co and Mn(B) couple ferromagnetically. Hence, the antiparallel alignment of the Mn(C) moment is stable with respect to Mn(B) \textit{and} Co. On the other hand, the intra-sublattice interactions are negative, which leads to a destabilization of the parallel alignment of the moments on one sublattice. It should be noted that in the Al--Ga--In series the Mn(C)-Mn(C) interaction is reduced on the first shell, while it is increased on the second shell at $r=1$, where it becomes larger than the Co-Mn(C) inter-sublattice interaction.

For the Si--Ge--Sn series some differences to the previous results are notable. The most evident one is the much lower Mn(C)-Mn(C) interaction, but also the Co-Mn(B) interaction is significantly reduced. In particular, the Mn(C)-Mn(B) interaction is reduced by a factor of about 3, in very good agreement with the reduction of the Mn(B) moment. This indicates a strong direct exchange interaction, which is feasible because of the small Mn(B)-Mn(C) distance of typically 2.53\,\AA{}. It is remarkable that the Co-Mn(B) interactions are even slightly increased with repect to the Al--Ga--In series, although the site-resolved magnetic moments are systematically lower. The additional loosely bound \textit{sp} electron augments the direct exchange coupling here. Finally, the intra-sublattice interaction of Mn(B)-Mn(B) is found to be positive in all three compounds on the first shell, but the other intra-sublattice parameters are still negative.

Mn$_2$CoSb is special in this respect, since it is a ferromagnet with a small positive magnetic moment on the Mn(C) site. Accordingly, the Co-Mn(C) and Mn(C)-Mn(B) interactions are positive, and their values are in reasonable agreement with the reduction of the Mn(C) moment. In contrast, the Co-Mn(B) interaction is still large and is with the exception of Mn$_2$CoSi the largest one among all discussed compounds. The Mn(B)-Mn(B) interaction is negative again on the first and second shells. Such a periodicity with respect to the valence electron count of the system has been predicted by \c{S}a\c{s}io\~{g}lu for some full Heusler compounds and occurs in the presence of indirect exchange interactions mediated by the conduction electrons \cite{Sasioglu08}.

\begin{table}[t]
\caption{\label{curie} Curie temperatures $T_\mathrm{C}^{\mathrm{MFA}}$ calculated in the mean-field approximation.}
\begin{indented}
\item[]\begin{tabular}{@{}l l l l l l l l l l l l l}
\br
Mn$_2$Co\textit{Z}	&	&			Al	& Ga	& In	& 	& Si	& Ge 	& Sn	& 	& Sb 	\\ \mr
$T_\mathrm{C}^{\mathrm{MFA}}$ (K)	&	& 	890	& 886	& 845	&	& 578	& 579	& 536	&	& 567	\\
\br
\end{tabular}
\end{indented}
\end{table}

From the exchange coupling parameters described above we calculated the Curie temperatures within the mean field approximation (see Table \ref{curie}). The series Al--Ga--In has surprisingly high values of more than 800\,K, even reaching almost 900\,K for Mn$_2$CoAl. For the Si--Ge--Sn series we found moderate values between 500 and 600\,K. The Curie temperature of Mn$_2$CoSb is similar as for the Si--Ge--Sn series. This is surprising at a first glance, because the Mn(C)-Mn(B) exchange interaction is so small here. It can be understood if we neglect all interactions but Mn(C)-Mn(B) and Co-Mn(B). In this case, $J_0^{\mu\nu}$ becomes a singular $3 \times 3$ matrix with two nonzero eigenvalues, which have the form of a root mean square of the Co-Mn(B) interaction and the Mn(C)-Mn(B) interaction. Obviously, if one interaction is significantly larger than the other (as, e.g., in Mn$_2$CoSn), then the eigenvalues will be dominated by the larger interaction. This and the increased Co-Mn(B) exchange interaction explain the unexpected behaviour.

However, what is most exciting about these results is the fact that Mn$_2$CoAl, Mn$_2$CoGa, and Mn$_2$CoIn have the highest Curie temperature among all ferrimagnetic intermetallic compounds reported to date. The Curie temperature decreases from one \textit{Z} group to another, although the total moment increases. A behaviour like this is unique for the Mn$_2$ based generalized Heusler compounds with Hg$_2$CuTi structure. The Co$_2$- and Mn$_2$-based genuine Heusler compounds show a scaling of the Curie temperature roughly proportional to the total moment upon change of the \textit{Z} element \cite{Kuebler07, Meinert10}. 

Naturally, the question about the accuracy of our Curie temperature calculation arises here. For the Mn$_2$Co\textit{Z} series only few data are available. Lakshmi \textit{et al.} reported $T_\mathrm{C} = 605$\,K for disordered Mn$_2$CoSn \cite{Lakshmi05}. Dai \textit{et al.} reported 485\,K for disordered Mn$_2$CoSb \cite{Dai06}. Hence, the $T_\mathrm{C}^{\mathrm{MFA}}$ value underestimates the measured value in Mn$_2$CoSn and overestimates it for Mn$_2$CoSb, so no systematic trend can be stated here. It is \textit{a priori} not clear which type of disorder can increase or decrease the Curie temperature, since the exchange interactions are highly site specific and quite complex. However, the calculated values reproduce the measured data within $\pm100$\,K. For comparison with Heusler compounds we also calculated the Curie temperatures of some compounds with available experimental data. The calculated (experimental) values are: Co$_2$MnSi 1049\,K (985\,K)\cite{Webster71}, Co$_2$TiSn 383\,K (355\,K)\cite{Majumdar05}, Mn$_2$VAl 605\,K (760\,K)\cite{Jiang01} and Mn$_2$VGa 560\,K (783\,K)\cite{Kumar08}. We overestimate the correct values slightly in the two Co$_2$-based Heusler compounds, but the values for the two Mn$_2$-based compounds are underestimated by 25\,\% or 150 to 200\,K. It is not clear where this deviation stems from; two possible sources should be considered. On the one hand, the magnetic ground state itself could be incorrect. For example, correlation effects (via an LDA+U treatment) can significantly enhance the Curie temperature \cite{Thoene09}. On the other hand, the calculation of the $J_{ij}$ could be incorrect, due to the approximations made here. Liechtenstein's expression considers a long-wavelength approximation, thus it is less suited for materials with short-range magnetic order, which is important, e.g., in fcc Ni \cite{Antropov03, Antropov06}.

\begin{figure}[t]
\centering\includegraphics[scale=1]{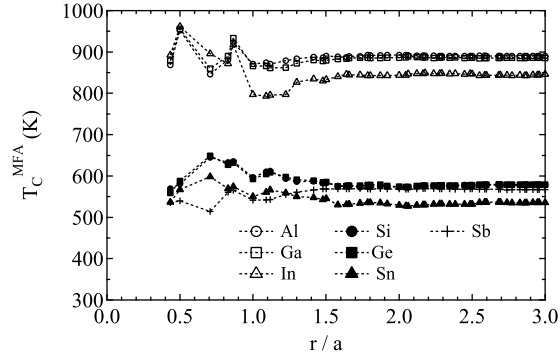}
\caption{\label{fig:Tc_vs_R}Curie temperatures $T_\mathrm{C}^{\mathrm{MFA}}$ as a function of the cluster radius taken into account.}
\end{figure}

In Figure \ref{fig:Tc_vs_R} we show the calculated Curie temperatures in dependence on the cluster radius taken into the summation in Equation (\ref{eq:multi}). As expected from the $J_{ij}$ plots in Figure \ref{fig:jxc}, $T_\mathrm{C}^{\mathrm{MFA}}$ is already determined by the nearest neighbor interactions in all compounds. Only weak changes are observed with increasing cluster radius and $T_\mathrm{C}^{\mathrm{MFA}}$ is well converged at $r = 1.5\,a$. This plot helps us to identify the origin of the reduced Curie temperatures of Mn$_2$CoIn and Mn$_2$CoSn, which is apparently not the same. For Mn$_2$CoIn we can assign the jump at $r = a$ to the strong antiferromagnetic intra-sublattice interaction of Mn(C)-Mn(C). In Mn$_2$CoSn, the reduced ferromagnetic Mn(B)-Mn(B) intra-sublattice interaction on the third neighbor shell at $r = 0.707\,a$ is responsible for the reduction.

In order to shed some more light on the character of the exchange interactions and their dependence on the site specific magnetic moments, we calculated the ground states and exchange coupling parameters for Mn$_2$CoGe in the range of $a = 5.60 \dots 5.95$\,\AA{}. Thereby we can separate the influence of the \textit{Z} valence electron count and the binding energy from geometric effects. The compound is a ferrimagnetic half-metal over the whole range, so we can expect minimal band structure effects on the calculations.

\begin{figure}[t]
\centering\includegraphics[scale=1]{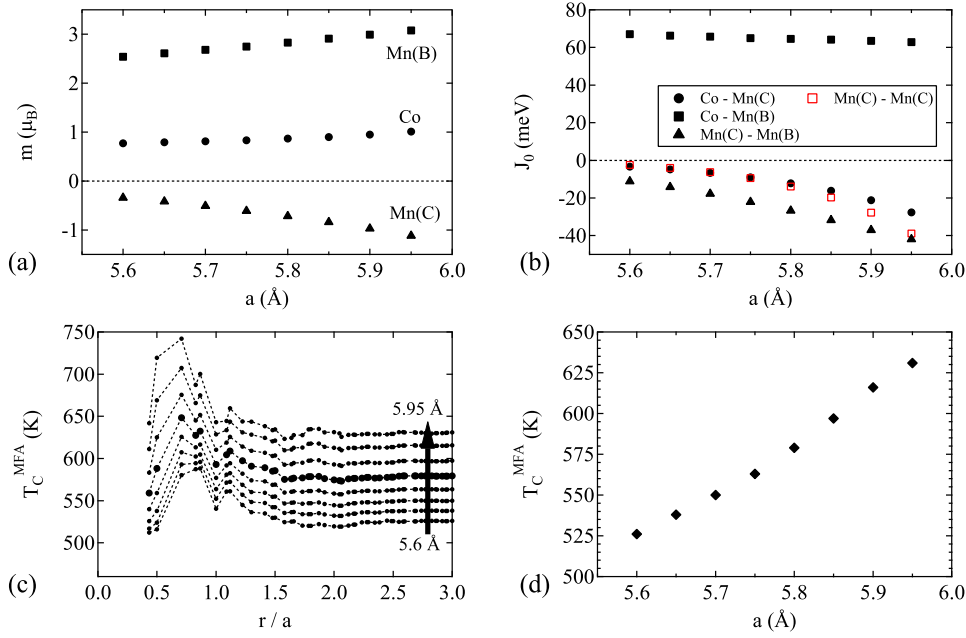}
\caption{\label{fig:Mn2CoGe}Lattice parameter dependencies in Mn$_2$CoGe. (a): Site resolved magnetic moments. (b): $J_0^{\mu \nu}$ contributions. (c): Curie temperatures $T_\mathrm{C}^{\mathrm{MFA}}$ in dependence on the cluster radius. (d): Curie temperatures $T_\mathrm{C}^{\mathrm{MFA}}$. }
\end{figure}

On the other hand, the site resolved magnetic moments change considerably with the lattice parameter (Figure \ref{fig:Mn2CoGe} (a)). Their absolute values increase with increasing lattice parameter as already explained above. All moments vary approximately linearly such that the total moment remains at 3\,$\mu_\mathrm{B}$\,/\,f.u. The moment of Mn(C) changes within the investigated range by more than a factor of three, from -0.34 to -1.12\,$\mu_\mathrm{B}$. The compensation comes mostly from the Mn(B) site, and the Co moment remains fairly constant.

To display the exchange interactions in a compact form, we show the relevant contributions to the $J_0^{\mu \nu}$ matrix (Equation (\ref{eq:multi})) in Figure \ref{fig:Mn2CoGe} (b). The Co-Mn(B) interaction sum is nearly constant, although the magnetic moments increase. The nearest neighbor interaction remains nearly constant, but the weak longer-ranging interaction is significantly decreased and accounts for the decrease in the interaction sum. The constant nearest-neighbor interaction is a result of two opposing processes, namely the increase of the moments and the reduction of exchange efficiency due to longer interatomic distances. In contrast, the interactions involving Mn(C) change considerably with the interatomic distance. The Mn(C)-Mn(B) exchange interactions increase by a factor of four, in agreement with the product $m_\mathrm{Mn(B)} \cdot m_\mathrm{Mn(C)}$. Further, the Co-Mn(C) interaction increases more than linearly with the lattice parameter, but the interaction is presumably indirect and no simple dependence is obvious. All these interactions lead to an increase of the Curie temperature with the lattice parameter. In contrast, the antiferromagnetic Mn(C)-Mn(C) exchange interaction counteracts the ferrimagnetic order in the compound and reduces the Curie temperature. This influence is, however, negligible at small lattice parameter, but becomes quite large at the highest values, even compensating the Mn(C)-Mn(B) interaction. Notably, the Mn(C)-Mn(C) interaction is entirely governed by the nearest neighbor interaction and depends approximately on $m_\mathrm{Mn(C)}^2$.

Figure \ref{fig:Mn2CoGe} (c) displays the Curie temperature in dependence on the cluster radius. The general features of the exchange interactions are the same for all lattice parameters considered. However, there are some subtle differences on the second and third shells at $r = 0.5\,a$ and $r=0.707\,a$, respectively. The change on the second shell can be traced back to the increased Co-Mn(C) interaction. Relative to the second shell, the contribution of the third shell is reduced. This arises from the increased antiferromagnetic Mn(C)-Mn(C) interaction discussed above. For clarity, Figure \ref{fig:Mn2CoGe} (d) shows that the resulting Curie temperature $T_\mathrm{C}^{\mathrm{MFA}}$ increases from 526\,K to 631\,K with increasing lattice parameter.

In terms of a pressure dependence, the Curie temperature of Mn$_2$CoGe is thus predicted to decrease upon hydrostatic pressure,  i.e., $\mathrm{d}T_\mathrm{C} / \mathrm{d} p < 0$. This situation is very different from that in Heusler compounds, where usually $\mathrm{d}T_\mathrm{C} / \mathrm{d} p > 0$ is found. However, it is in agreement with Castelliz' \cite{Castelliz55} and Kanomata's \cite{Kanomata87} empirical interaction curves. They propose a negative pressure coefficient of $T_\mathrm{C}$ for short Mn-Mn distances as in hexagonal MnAs or MnSb, but a positive coefficient at larger distances as in the Heusler compounds \textit{X}$_2$Mn\textit{Z}. \textit{Ab initio} calculations by Yamada \textit{et al.} on hexagonal MnAs \cite{Yamada02} and by  \c{S}a\c{s}io\~{g}lu \textit{et al.} on the Heusler compound Ni$_2$MnSn \cite{Sasioglu05_2} are in agreement with the experimentally observed pressure dependencies. Recently, we have also calculated a positive pressure coefficient of $T_\mathrm{C}$ in the (hypothetical) Mn$_2$Ti\textit{Z} Heusler compounds \cite{Meinert10}. The Mn-Mn distance in the Mn$_2$Co\textit{Z} compounds is even smaller than in the hexagonal MnAs or MnSb compounds, so a strong negative pressure dependence of $T_\mathrm{C}$ is in good agreement with the available experimental data.

Following from the lattice parameter dependence, the reduction of $T_\mathrm{C}^{\mathrm{MFA}}$ in Mn$_2$CoIn and Mn$_2$CoSn (which have the largest lattice parameters within their groups) can be ascribed to a binding energy effect due to the high-lying valence states in In and Sn. 

\section{Conclusion}

We have performed \textit{ab initio} band structure calculations with the full potential Korringa-Kohn-Rostoker method on the Mn$_2$Co\textit{Z} compounds with the Hg$_2$CuTi structure. The exchange interaction parameters obtained from the calculations are found to be governed by the Co-Mn(C) exchange, which is of direct nature. In the case of \textit{Z} = Al, Ga, In, the Mn(C)-Mn(C) interaction is the dominating one, which is direct as well. The indirect, long-ranged interactions are exponentially damped and thus weak, and the intra-sublattice interactions are mostly antiferromagnetic. Curie temperatures calculated within the mean-field approximation are in reasonable agreement with experimental data for Mn$_2$CoSn and Mn$_2$CoSb. The Curie temperatures show an anomalous dependence on the total moment, which is different from the full Heusler compounds. For Mn$_2$CoAl we predict an exceptionally high Curie temperature of 890\,K, although the total moment of the compound is only 2\,$\mu_\mathrm{B}$\,/\,f.u. The dependence of the exchange parameters on the lattice parameter in Mn$_2$CoGe suggests a negative pressure dependence of $T_\mathrm{C}$ in the Mn$_2$Co\textit{Z} compounds, which originates from the exchange interactions of Mn(C)-Mn(B) and Co-Mn(C).

\section*{Acknowledgements}
The authors gratefully acknowledge financial support from Bundesministerium f\"ur Bildung und Forschung (BMBF) and Deutsche Forschungsgemeinschaft (DFG).

\end{document}